# Machine Learning Integrated Near-Infrared Surface-Enhanced Raman Spectroscopy for Accurate Strain-Level Virus Identification

Na Zhang[1,†,*], Ziyang Wang[2,†], Xielin Wang[2], Gabriel A. Vázquez-Lizardi[3], Paula Piñeiro Varela[4], Dorleta Jimenez de Aberasturi[4,5], David E. Sanchez[6], Nestor Perea-Lopez[1], Samuel Lin[2], Ryeanne Ricker[7], Edgar Dimitrov[1], Alexander J. Sredenschek[1], Kalana D. Halanayake[3], Yin-Ting Yeh[1], Julian A. Mintz[1], Jiarong Ye[8], Sharon Xiaolei Huang[8], Huaguang Lu[9], Elodie Ghedin[7], Danielle Reifsnyder Hickey[3,6,10], Luis M. Liz-Marzán[4,5,11], Shengxi Huang[2,12,*], Mauricio Terrones[1,3,4,11,13,*]

[1] Department of Physics, The Pennsylvania State University, University Park, Pennsylvania 16802, USA

[2] Department of Electrical and Computer Engineering, Rice University, Houston, Texas 77005, USA

[3] Department of Chemistry, The Pennsylvania State University, University Park, Pennsylvania 16802, United States

[4] CIC biomaGUNE, Basque Research and Technology Alliance (BRTA), 20014 Donostia-San Sebastián, Spain

[5] Ikerbasque, Basque Foundation for Science, 48009 Bilbao, Spain

[6] Department of Materials Science and Engineering, The Pennsylvania State University, University Park, Pennsylvania 16802, USA

[7] Systems Genomics Section, Laboratory of Parasitic Diseases, National Institute of Allergy and Infectious Diseases, National Institutes of Health, Bethesda, MD 20894

[8] College of Information Sciences and Technology, The Pennsylvania State University, University Park, PA 16802

[9] Department of Veterinary and Biomedical Sciences, The Pennsylvania State University, University Park, PA 16802

[10] Materials Research Institute, The Pennsylvania State University, University Park, Pennsylvania 16802, USA

[11] Two-Dimensional Crystal Consortium, The Pennsylvania State University, University Park, Pennsylvania 16802, United States

[12] Rice Advanced Materials Institute, Rice University, Houston, Texas 77005, USA

[13] Center for 2-Dimensional and Layered Materials, The Pennsylvania State University, University Park, Pennsylvania 16802, United States

[†] These authors contributed equally.
*Corresponding authors: nmz5113@psu.edu, shengxi.huang@rice.edu, mut11@psu.edu (M.T.)

## Abstract

Strain-level identification of viruses is critical for effective public health responses to potential outbreaks, yet current diagnostic methods often lack the necessary speed or sensitivity. Surface-enhanced Raman spectroscopy (SERS) offers great potential for fast and precise virus clarification through the unique vibrational fingerprints of biological components. However, existing protocols typically operate outside of the tissue's transparent near-infrared (NIR) window, and are further limited by the intrinsic complexity of clinical viral samples, which complicates spectral analysis and recognition. Here, we report an artificial intelligence (AI)-empowered NIR-SERS platform that integrates machine learning with a rationally designed hybrid substrate: gold nanostars (AuNSt) coupled with gold-coated carbon nanotube arrays (AuCNT). This architecture generates highly localized plasmonic hot spots resonant tuned to NIR excitation, as confirmed by electron energy-loss spectroscopy (EELS), enabling effective signal amplification from viral components. Our system and protocols provide accurate classification of respiratory viruses, including influenza viruses and coronaviruses, not only at the type and subtype levels, but also the more challenging strain level. This approach overcomes the plasmonic mismatch in conventional SERS and the lack of generalizability in AI-driven diagnostics. It shows promise for enhancing rapid virus detection and identification of novel strains and outbreak response capabilities, thus potentially addressing critical challenges in global public health preparedness.

## Introduction

Viruses are prone to frequent mutations, giving rise to new variants and strains that can differ in transmissibility, virulence, and resistance to existing treatments or vaccines [1,2]. Rapid and accurate identification of viruses at the strain level is crucial for effective epidemiological surveillance, timely therapeutic decisions, and informed public health interventions[3]. Traditional diagnostic methods, such as polymerase chain reaction (PCR) and antigen tests, have inherent limitations: For example, PCR requires relatively long processing times and specialized personnel[4], while antigen tests suffer from low sensitivity, especially in the early stages of infection[5]. These challenges underscore the urgent need for alternative detection platforms that are not only fast and label-free but also capable of distinguishing viruses with high sensitivity and resolution down to the strain level.

Surface-enhanced Raman spectroscopy (SERS) provides a label-free approach for molecular identification with both rapid analysis and high sensitivity through the distinct vibrational fingerprints of biological components, a combination of advantages not typically achieved by other detection methods.[6-8] The remarkable sensitivity of SERS originates from localized surface plasmon resonance (LSPR) in metallic nanostructures, which generates highly confined electromagnetic fields capable of amplifying Raman signals by several orders of magnitude, down to the single-molecule level[9]. Nevertheless, a key challenge in Raman-based bioanalysis is strong fluorescence and potential photodamage associated with laser excitation in the visible range[10,11]. Near-infrared (NIR) excitation,- known as the "tissue diagnostic window", mitigates these issues while enabling deeper penetration into biological samples compared to visible lasers. Although 785 nm excitation has been widely used in biomedical Raman applications[12], most SERS substrates are still optimized for resonance in the visible wavelengths[13,14]. This plasmonic mismatch leads to diminished enhancement efficiency, and this problem is further compounded by the fact that light scattering are inversely proportional to the fourth power of the laser wavelength[15], making signal amplification under NIR conditions particularly crucial.

Moreover, clinical viral samples possess intrinsic complexity, which complicates the extraction and interpretation of spectral fingerprints. These challenges underscore the need for advanced data-processing strategies to enable reliable and automated spectral analysis for virus identification. Herein, we propose a strategy that integrates artificial intelligence (AI) with NIR resonant SERS, employing a rationally designed hybrid substrate composed of gold nanostars (AuNSt) and gold-coated carbon nanotube (AuCNT) arrays engineered for optimal plasmonic resonance in the NIR range. This platform significantly amplifies the Raman signals of viruses excited at 785 nm. When coupled with machine learning algorithms, the NIR-active SERS substrate enables highly accurate classification of viruses not only at the type and subtype levels but also the more challenging strain level. Our result highlights the potential of AI-empowered NIR-SERS platform as a powerful platform for rapid and precise detection and identification of viral pathogens during outbreak surveillance and diagnostics.

## Results

**Hybrid substrate for NIR SERS enhancement**

We previously developed a virus-capture and detection platform, which utilizes vertically aligned nitrogen-doped carbon nanotube arrays coated with rough gold films (AuCNT), which enables size-based virus capture, enrichment and in-situ Raman characterization[13,14]. However, the signal-to-noise ratio of the Raman spectra was not sufficient for reliable strain level classification due to the plasmonic mismatch. To address this challenge and maximize signal enhancement of viruses under the 785 nm excitation wavelength, we leveraged the tunable plasmonic property of AuNSt. Compared to the gold nanospheres (AuNP) with a restricted LSPR tunability (~520–580 nm), AuNSt can be shifted from the red to the NIR region by controlling the aspect ratio of their spikes[16]. Gold nanocubes, though offering a redshift due to their geometry, provide limited field enhancement primarily at their vertices[17]. The multiple sharp spikes of AuNSt generate a strong localized electromagnetic field at their tips, making them ideal for SERS enhancement in the NIR spectral range.

Based on this rationale, we designed a hybrid SERS substrate by incorporating AuNSt into the existing AuCNT platform. As illustrated in Figure 1a, the vertically aligned CNT arrays were synthesized on a Si substrate by chemical vapor deposition (CVD), using patterned iron (Fe) particles as catalysts[18]. Gold was subsequently deposited and annealed to form discontinuous rough gold films and nanoparticles on the CNT surface (Figure S1). Separately, the AuNSt with an average (tip-to-tip) diameter of 50 nm were synthesized via a seed mediated growth and characterized using bright-field transmission electron microscopy (TEM; Figure S2a)[19]. The AuNSt were then deposited onto the AuCNT platform via drop casting. Representative scanning electron microscopy (SEM) and TEM images show the physical attachment of AuNSt to the

AuCNT surface (Figure 1b, c). The coupling between AuNSt tips and the underlying rough gold film is expected to generate abundant plasmonic "hot spots" and shift the plasmon resonance into the NIR region, thereby significantly improving the SERS effect while maintaining the biocompatibility and viral capture efficiency of the AuCNT platform.

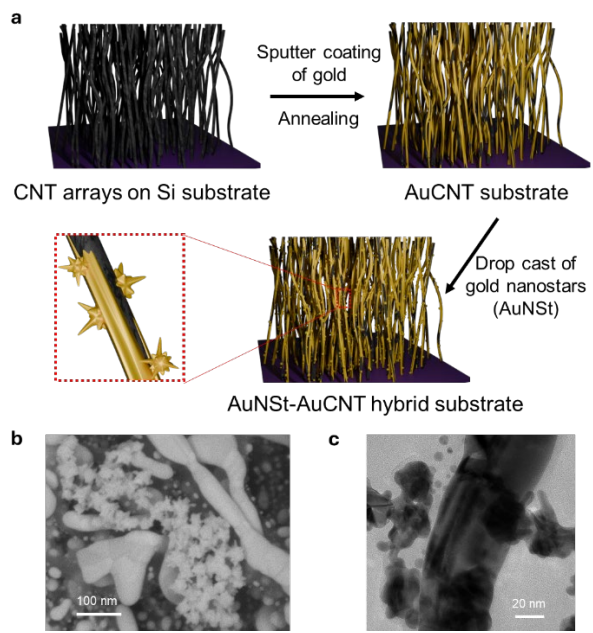

**Figure 1.** Scheme depicting the AuNSt-AuCNT hybrid substrate for SERS enhancement. Schematic illustration of (**a**) the hybrid SERS substrate preparation and the AuNSt-AuCNT hybrid structure. Scanning electron microscopy ((**b**) SEM) and transmission electron microscopy ((**c**) TEM) images of AuNSt-AuCNT hybrid substrate. AuCNT: gold carbon nanotubes; AuNSt: gold nanostars; Si: silicon.

**NIR plasmon resonance of the AuNSt-AuCNT hybrid substrate**

To validate the plasmonic behavior and confirm the NIR resonance of the hybrid substrate, we performed a combination of optical characterization and spatially resolved electron energy loss spectral imaging (EELS). Optical properties were assessed using UV-vis spectroscopy and reflection absorption spectroscopy. Whereas the UV-vis spectrum of the colloidal gold nanospheres (AuNP) solution shows a plasmon band centered at 535 nm, the AuNSt solution exhibits a broad extinction band at around 700 nm, accompanied by a weaker shoulder near 530 nm (Figure S2b). These features correspond to the plasmon resonance originating from the nanostar branches and core, respectively[16]. The 3D finite-difference time-domain (FDTD) method was utilized to simulate the extinction of AuNP and AuNSt solutions. AuNP were modeled as simple spheres and the AuNSt were represented by a core sphere surrounded by 20 truncated cones (referred to as cones below) with geometrical parameters derived from statistical analysis of AuNSt in TEM images(Figure S2c-f). By sweeping the cone height and tip radius (see *Supporting Information* and Figure S2g), an optimal structure model consisting of a 7 nm core radius, 20 nm cone height and averaged top radius between 1–2 nm, yielded the best match to the experimental UV-vis spectrum (Figure S2h). The simulated electromagnetic field distribution of AuNSt highlights strong electromagnetic field enhancement at the nanostar tips (Figure S2i), corresponding to the anticipated SERS "hot spots".

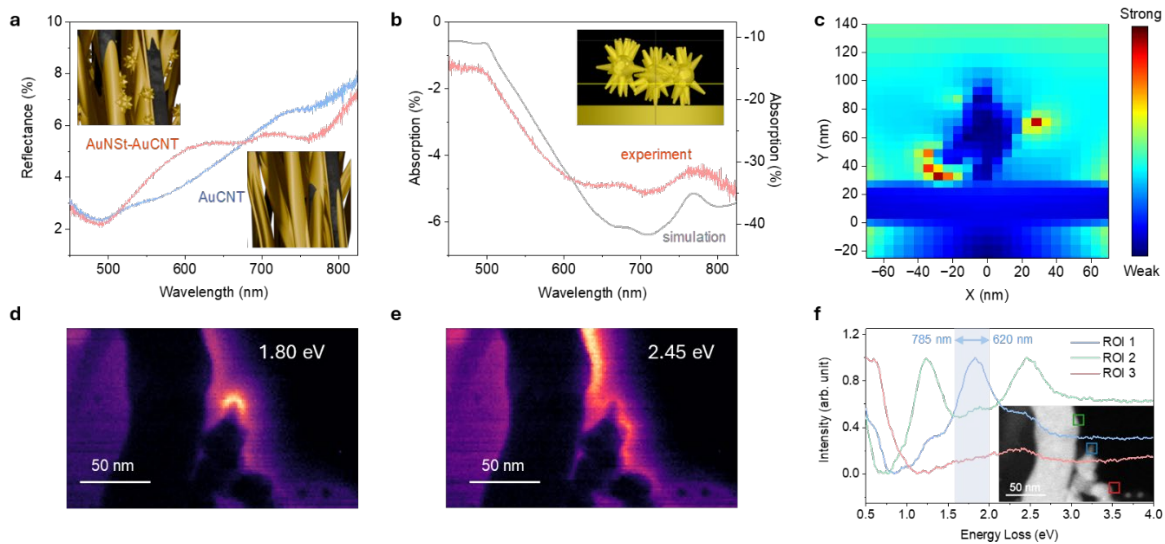

**Figure 2.** NIR resonance of the AuNSt-AuCNT hybrid substrate. **a**. Reflectance spectra of the AuNSt-AuCNT hybrid substrate and standard AuCNT substrate. **b**. FDTD simulated absorption of the AuNSt-AuCNT hybrid substrate compared with the experimentally obtained one. **c**. The simulated local electromagnetic field distribution of a single AuNSt on AuCNT hybrid substrate. Plasmon maps obtained experimentally from the integrated EELS signal over the energy range: **(d)** $1.80 \pm 0.1$ eV ($688 \pm 70$ nm) **(e)** $2.45 \pm 0.1$ eV ($506 \pm 50$ nm). **f**. EELS spectra obtained at different regions of interest (ROIs, marked by boxes with colors corresponding to the spectra).

The plasmon resonance features of the non-transparent AuNSt-AuCNT hybrid substrate were measured using reflectance spectroscopy[20]. As shown in Figure 2a, there is a dominant reflectance dip centered at around 765 nm in the reflectance spectra of the AuNSt-AuCNT hybrid substrate, which is absent in that of the AuCNT substrate. This NIR resonance, aligning with the 785 nm excitation wavelength, indicates strong coupling between AuNSt and AuCNT. FDTD simulations of the hybrid structure, incorporating various AuNSt configurations on AuCNT (see *Supporting Information* and Figure S3), resulted in absorption spectra closely matching experimental results, featuring a strong plasmon band at around 765 nm together with a similar shoulder at around 680 nm, as observed experimentally (Figure 2b). The corresponding simulated electromagnetic field distribution of a single AuNSt on AuCNT system (Figure 2c) confirms the strongest field enhancement is localized at the AuNSt tips and the junction between the tips of the AuNSt branches and the underlying gold film, supporting efficient NIR SERS performance.

While the optical characterization and FDTD simulations provide ensemble-averaged information about the plasmonic behavior, they are inherently limited in spatial resolution and cannot directly resolve where plasmonic enhancement occurs within complex nanostructures. To overcome this limitation, electron energy-loss spectral imaging (EELS) was performed to map localized surface plasmon resonances (LSPRs) with nanometer-scale precision[21]. The loss intensity map obtained from the integrated EELS signal over the energy range of 1.70- 1.90 eV (Figure 2d) clearly demonstrate the pronounced field enhancement on the far tip of a AuNSt that is adjacent to the gold film, as identified in the corresponding scanning TEM (STEM) image (inset in Figure 2f). This localized response confirms the capability of the AuNSt-AuCNT hybrid structure to generate significant field enhancement in the NIR range. In contrast, the EELS maps centered at around 2.45 eV (Figure 2e) displays plasmon excitation across both the AuNSt and gold film surface, indicating comparable electromagnetic enhancements in the visible range. The EEL spectra extracted from regions of interest (ROIs; Figure 2f) further support our interpretation: ROI 1 (where the AuNSt tips and Au film are in proximity) exhibits a dominant peak at around 1.80 eV, whereas ROI 2 (on the gold film but not adjacent to a AuNSt) shows a weaker but discernible peak at the same energy. Both regions display a strong plasmon resonance around 2.45 eV. As a comparison, ROI 3 (an unstructured gold region that is not adjacent to the AuNSt) shows only a broad, featureless background over the energy range of 1.10- 2.60 eV. These findings are in good agreement with optical characterization and FDTD calculations, thus collectively confirming the AuNSt-AuCNT hybrid substrate supports an intense NIR plasmon resonance, which is essential for enhancing Raman scattering under the 785 nm excitation wavelength.

**NIR SERS performance of the AuNSt-AuCNT hybrid substrate**

To evaluate the NIR SERS performance of the AuNSt-AuCNT hybrid substrate, we employed zinc phthalocyanine (ZnPc) as a probe molecule. ZnPc has a HOMO-LUMO gap of 1.4 eV[22], which falls within the NIR region, making it suitable for assessing enhancement under the 785 nm laser excitation. For comparison, ZnPc was deposited via vacuum thermal deposition onto four different substrates: bare thermally oxidized silicon ($SiO_2$/Si), $SiO_2$/Si with sputtered gold film, AuCNT, and the AuNSt–AuCNT hybrid substrate (see Figure 3a). As shown in Figure 3b, ZnPc on the gold film exhibited significantly enhanced Raman signals compared to bare $SiO_2$/ Si, confirming plasmonic enhancement under the 785 nm excitation. However, ZnPc deposited on the AuCNT substrate showed negligible Raman response, despite sharing the same gold film deposition as the planar substrate. This is attributed to the vertical orientation and depth of the CNT arrays, which limit the amount of ZnPc on the confocal plane of the Raman laser. Interestingly, the AuNSt–AuCNT hybrid substrate demonstrated strong Raman signals for ZnPc despite the same molecular loading limitation, suggesting substantial electromagnetic enhancement on the AuNSt–AuCNT platform. The enhancement factor (EF) for the ZnPc Raman mode at around 1512 cm$^{-1}$ on AuNSt–AuCNT hybrid substrate relative to Si substrate was calculated to be around 2000 (see Figure S4 and *Supporting Information*).

Laser excitation dependence further supports the NIR resonance enhancement of the AuNSt–AuCNT hybrid substrate. Under 633 nm excitation with similar laser power, Raman peaks of ZnPc were significantly weaker (Figure 3c), consistent with the off-resonance condition of the substrate's plasmonic response and the electronic transitions of ZnPc from the excitation wavelength. To demonstrate that the NIR enhancement is not limited to molecules resonant with the laser line, we tested copper phthalocyanine (CuPc) and crystal violet (CV) molecules, which have a HOMO-LUMO gap of 1.7 eV and 1.9 eV, respectively, higher than the 785 nm laser energy (1.58 eV) (see Figure S5)[23]. Under 785 nm laser excitation, the EF for the CuPc Raman mode near 1531 cm$^{-1}$ was estimated to be around 560, while the EF for the CV Raman mode near 1172 cm$^{-1}$ was around 630. These values are likely underestimated, as the corresponding signals on Si were undetectable and the intensity was approximated using the noise level, which may in fact be even lower. Although lower than for ZnPc, such strong enhancements for non-resonant molecules underscore the versatility of the NIR SERS substrate. Consistently, the strongest enhancements were observed for ZnPc molecules whose absorption overlaps with the NIR, highlighting the importance of resonance matching for optimal SERS performance.

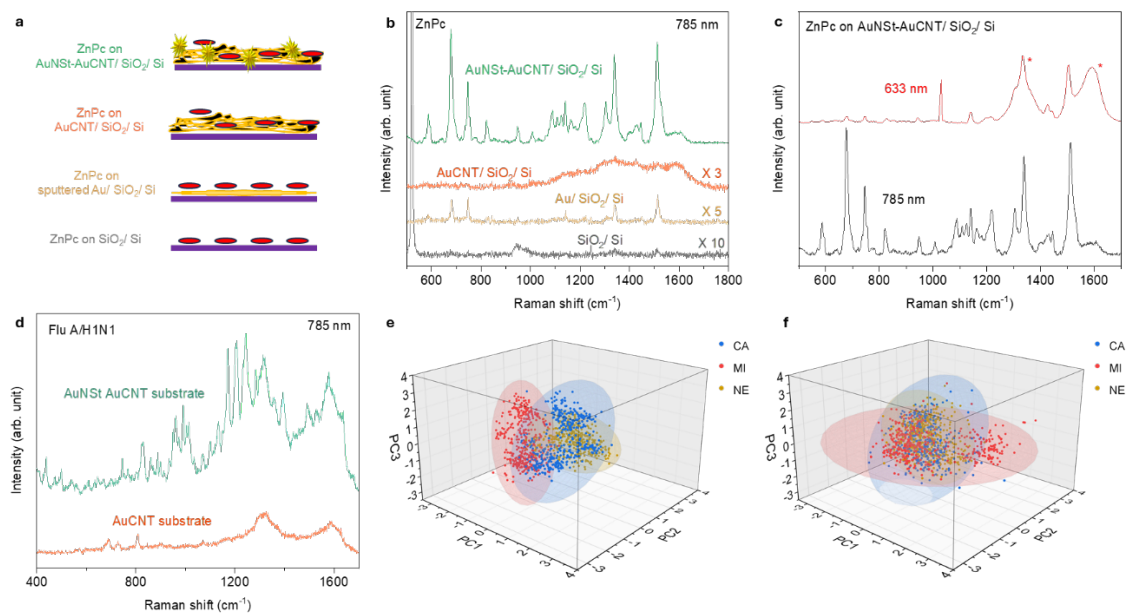

**Figure 3.** NIR SERS performance of the AuNSt-AuCNT hybrid substrate. **a**. Overlaid configurations of different SERS substrates for ZnPc molecules. **b**. Corresponding enhanced Raman spectra of ZnPc on these substrates. **c**. The enhanced Raman spectra of ZnPc on the AuNSt-AuCNT hybrid substrate under excitation lasers of 785 nm (black) and 633 nm (red). The asterisk labels Raman peaks of CNT. **d**. Raman spectra of Flu influenza A, H1N1 viruses (CA: A/California/07/2009) obtained on AuNSt-AuCNT hybrid substrate and normal AuCNT substrate. PCA plots of Raman spectra collected from different H1N1 strains on (**e**) AuNSt-AuCNT hybrid substrate and (**f**) the standard AuCNT substrate. CA (blue): A/California/07/2009; MI (red): A/Michigan/45/2015; NE (yellow): A/Nebraska/14/2019.

To evaluate the NIR SERS performance of the AuNSt-AuCNT hybrid substrate for viral samples, we collected Raman spectra from human respiratory viruses, including an avian respiratory virus (IBV), influenza B (Flu B), and influenza A (Flu A) with two subtypes, H1N1 and H3N2, influenza B (Flu B), and avian infectious bronchitis viruses (IBV), across multiple strains (see *Methods*).

The rational design of AuNSt-AuCNT hybrid substrates with nanostructural features comparable in size to viruses not only facilitates efficient coupling between the analytes and plasmonic hotspots but also improves biocompatibility, which is essential for accurate viral detection and characterization[13]. Figure S6 demonstrates the Raman spectra of Flu A/H3N2 viruses acquired from four different locations on the AuNSt-AuCNT hybrid substrate. The variation of spectra between these spots reflects the heterogeneity of viral structure and the spatially localized nature of plasmonic "hot spots", underscoring the importance of collecting richer Raman features from the viruses to guarantee precise recognition of different viruses. Compared to the AuCNT substrate, the AuNSt–AuCNT hybrid platform produced richer Raman features and superior signal-to-noise ratios for virus detection. As shown in Figure 3d, the Raman spectrum of Flu A/H1N1 viruses obtained on the AuNSt-AuCNT hybrid substrate has significantly more distinct peaks and better signal-to-noise ratio (231, taking the peak at 1247 $cm^{-1}$) than that acquired on the AuCNT substrate (45, taking the peak at 1247 $cm^{-1}$), highlighting the impressive SERS performance of the AuNSt-AuCNT hybrid substrate in the NIR region.

To illustrate the distribution and separability of Raman spectral data collected on the two types of substrates, we applied principal component analysis (PCA) to visualize the distribution of spectra obtained from three Flu A/H1N1 strains. Each strain was represented by n = 400 spectra, and each Raman spectrum was projected into a three-dimensional PCA space, as shown in Figure 3e. In this representation, each point corresponds to a single Raman spectrum. The PCA plot reveals that spectra acquired on the AuNSt-AuCNT hybrid substrate exhibit clear clustering and strong separability among the three viral strains. This demonstrates that the hybrid substrate enhances spectral distinction between strains. In contrast, the overlapping clusters observed for the AuCNT substrate reflect reduced discriminatory power (Figure 3f). To further validate the separability and explore the data structure in a non-linear embedding, we also applied t-distributed stochastic neighbor embedding (t-SNE) to the same datasets (Fig. S7). Additional PCA plots demonstrating the spectral separability across different viral types (Flu A, Flu B, and IBV), subtypes (H1N1 vs. H3N2), and strains of H3N2 are provided in Figure S8. These expanded analyses support the generalizability of the substrate's enhancement across different viral classes.

**Machine learning guided virus identification**

In order to identify enable accurate viruses identification based on their Raman spectra with high accuracy in distinguishing between different virus types, subtypes, and strains, we employed several machine learning approaches, including one-dimensional convolutional neural network (1D CNN), linear support vector machine (SVM), and XGBoost, as described in previous studies[13,24,25]. Figure 4a presents representative Raman spectra from three levels of viral classification: virus types (IBV, Flu B, and Flu A), subtypes (Flu A H1N1 and H3N2), and strains (H3N2 strains in various sources of HI, AZ, and DE). The 1D CNN architecture for virus identification is demonstrated in Figure 4b. Input Raman spectra are processed through a series of convolutional layers that extract key spectral features, particularly Raman shifts corresponding to biochemical signatures, and are then mapped to predicted class labels. This process enables the model to learn discriminative patterns directly from raw spectra without manual feature engineering.

Figure 4c compares the performance of the three machine learning models in classifying virus type, subtype, and strain level. Model performance was evaluated using precision, recall, and F1-score[26]. Precision measures the proportion of true positive predictions among all predicted positives. Recall quantifies the proportion of true positives identified among all actual positives. The F1-score, the harmonic mean of precision and recall, provides a balanced evaluation that accounts for both false positives and false negatives. Among the models, the 1D CNN consistently achieved the highest performance across all virus classification tasks, with precision, recall, and F1-scores exceeding 0.97. Specifically, on the AuNSt-AuCNT hybrid substrate, the CNN achieved an F1-score of 0.98 for virus type, 0.97 for subtype, and up to 0.99 for H1N1 and 0.97 for H3N2 strain classification. These results reflect the CNN's superior ability to extract informative spectral features and generalize effectively across classification levels. The XGBoost classifier also performed well, achieving F1-scores above 0.94 across all classification tasks. In contrast, the linear SVM showed lower performance, particularly for virus type and subtype classification, with F1-scores of 0.91 and 0.92, respectively.

To assess the impact of substrate on classification performance, we trained machine learning models using spectra collected on the standard AuCNT substrate (Figure S9). A notable decline in performance was observed when compared to the hybrid substrate. For instance, the CNN achieved classification F1-scores of 0.89 for virus type, 0.81 for subtype, 0.81 for H1N1 strain, and 0.72 for H3N2 strain on the AuCNT substrate. The performance of SVM and XGBoost models also dropped substantially on this substrate, with the SVM reaching as low as 0.61 F1-score for H3N2 strain classification. These results reinforce the superior quality of spectra acquired on the AuNSt-AuCNT hybrid substrate.

Furthermore, a consistent trend was observed across all models and substrates: Classification accuracy decreases progressively from virus type to subtype and then to strain, reflecting the increasing similarity and classification difficulty among closely related viral classes (see Figure 4d and Figure S10). While the CNN maintained F1-scores above 0.97 across all tasks on the hybrid substrate, its performance on the AuCNT substrate declined more noticeably, with F1-scores dropping from 0.89 for virus type, to 0.81 subtype, and further to 0.81 for H1N1 and 0.72 for H3N2 strain classification. This highlights the importance of strong SERS signal enhancement for reliable virus differentiation, particularly at the strain level. Our findings confirm that NIR SERS using the AuNSt-AuCNT hybrid substrate enables more robust, accurate, and granular virus identification.

To interpret the classification decision from the CNN and identify the spectral regions most relevant to virus differentiation, we generated feature importance maps (Figures 4e and 4f) by computing gradients with respect to input Raman wavenumbers. The map derived from the AuNSt–AuCNT hybrid substrate shows clearer and more distinct spectral regions that contribute to the model's predictions, including 619, 693, 755, 784, 1048, 1138, 1424, and 1585 cm$^{-1}$. These correspond to well-defined, virus-specific Raman features (Figure S11)[13,27-29]. For example, the feature at 784 cm$^{-1}$ corresponds to RNA, capturing nucleic acid signatures important for strain-level discrimination. Peaks at 1048 and 1138 cm$^{-1}$ highlight phenylalanine and lipid modes, indicating combined protein and membrane contributions. At higher wavenumbers, the 1424 and 1585 cm$^{-1}$ bands represent tyrosine, CH$_2$/CH$_3$ bending from lipids, and RNA bases, together marking protein structure, envelope composition, and genomic content.

In contrast, the feature importance map from the AuCNT substrate appears sparser and less resolved, with only weak contributions around 998–1007 cm$^{-1}$, indicating a lack of rich biochemical information. These differences highlight that the AuNSt-AuCNT hybrid substrate not only improves the quality of raw Raman spectra but also enhances the interpretability of machine learning models. The discrimination between virus types, subtypes, and strains is further illustrated in Figures S12–S14. Overall, the CNN leverages a biochemically interpretable set of Raman markers, including proteins (amide I and III, phenylalanine, tyrosine), lipids, and nucleic acids (RNA), to achieve accurate viral classification. The hybrid substrate amplifies these spectral features, enabling the model to resolve subtle but meaningful molecular differences that are otherwise obscured on the standard AuCNT substrate.

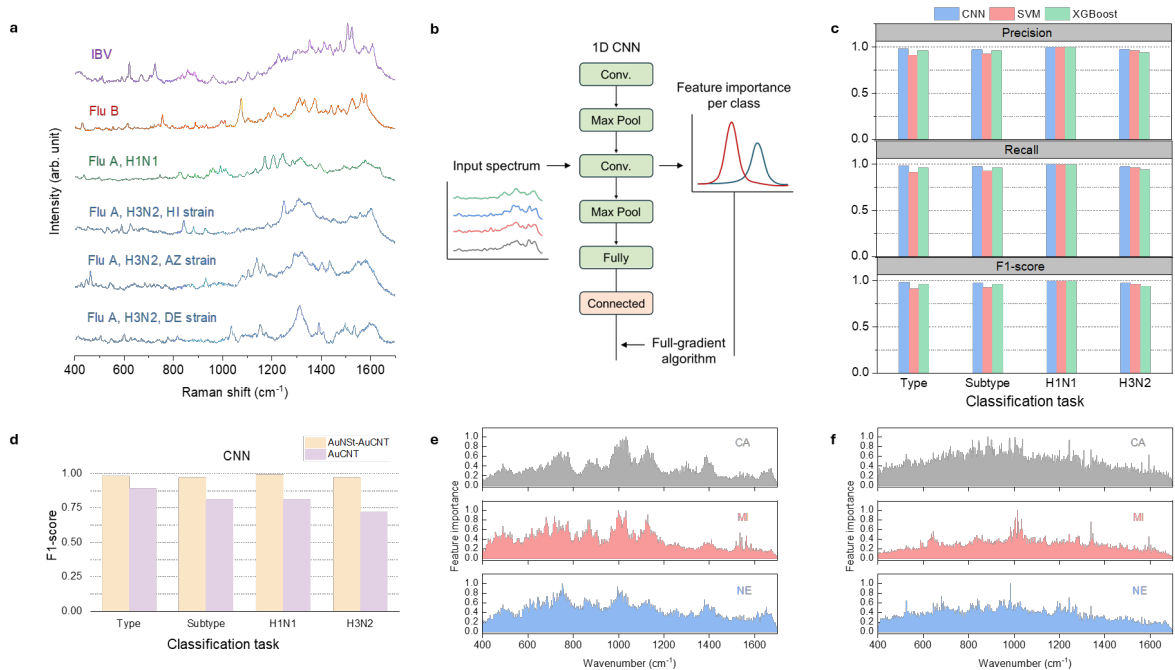

**Figure 4.** Machine learning guided virus identification. **a**. Representative Raman spectra of different virus types: IBV (infectious bronchitis virus), Flu B (B/Massachusetts/02/2012) and Flu A; virus subtypes: Flu A H1N1 (A/California/07/2009) and H3N2; virus strains: H3N2 HI (A/Hawaii/47/2014), AZ (A/Arizona/45/2018), and DE (A/Delaware/39/2019) strains. **b**. Schematic of the 1D CNN architecture used for virus identification and extraction of Raman feature important maps. **c**. Classification performance of CNN, SVM, and XGBoost models on 3 tasks (classification of virus type, subtype, and strain). **d**. Comparison of CNN classification performance using Raman spectra collected from the AuNSt-AuCNT hybrid substrate and normal AuCNT substrate. Feature importance maps of the CNN model using spectra from the AuNSt-

AuCNT hybrid substrate (**e**) and normal AuCNT substrate (**f**), indicating spectral regions contributing most to classification decisions.

## Discussion

In summary, we have developed an AI-empowered NIR SERS platform by integrating gold nanostars with gold-coated carbon nanotube arrays, enabling enhanced Raman signal amplification under 785 nm excitation. This design effectively addresses the fluorescence and signal loss challenges commonly existing in Raman-based bioanalysis. Combined with machine learning algorithms, the system achieved accurate classification of virus types, subtypes, and strains based on their improved spectral fingerprints.

This work establishes a rapid, label-free approach for high-resolution detection and identification of viruses, offering an effective tool for real-time viral pathogen surveillance and pandemic preparedness. Looking forward, further integration with microfluidic sampling and portable Raman systems could facilitate on-site diagnostics, while expansion to broader viral and bacterial panels may extend its utility in clinical and environmental monitoring. Moreover, our framework can be adapted to detect emerging novel virus strains or variants, contributing to proactive infectious disease management.

## Methods

**Synthesis of gold nanostars (AuNSt)**: AuNSt were synthesized using a seed-mediated growth method. To prepare the seed solution, 5 mL of 34 mM citrate was added to 95 mL of boiling 0.5 mM $HAuCl_4$ under vigorous stirring. After 15 minutes, the solution was cooled and stored at 4 °C. For the growth of 50 nm AuNSt, 2.5 mL of the citrate-stabilized seed solution was added to 50 mL of 0.25 mM $HAuCl_4$ containing 50 μL of 1 M HCl in a 20 mL glass vial at room temperature with moderate stirring. After that, 500 μL of 3 mM $AgNO_3$ and 250 μL of 100 mM ascorbic acid were added simultaneously, causing the solution to change from light red to greenish, indicating AuNSt formation. The resulting colloid was functionalized by adding 410 μL of 0.1 mM PEG-SH and stirred for 15 minutes. The product was then purified by centrifugation at 1190g for 25 minutes at 10 °C and redispersed in water.

**Synthesis of carbon nanotubes**: Patterned iron catalyst thin films were fabricated on single-side polished prime silicon wafers. The wafers were thoroughly cleaned with acetone, isopropyl alcohol, and ultrapure water. A double layer of photoresists was then spin-coated and patterned via photolithography. Iron was subsequently deposited using an e-beam evaporator under high vacuum ($10^{-7}$ Torr) at a rate of 0.1 nm/s, targeting thicknesses of 5 nm. The actual measured thickness was confirmed by atomic force microscopy. The wafers were diced into 14 mm by 14 mm individual dies. These iron-patterned substrates were then used for the synthesis of nitrogen-doped multi-walled carbon nanotubes (N-MWCNTs) via an aerosol-assisted chemical vapor deposition (AACVD) setup. This system comprised an ultrasonic nebulizer, two series-connected tube furnaces, and a waste trap containing isopropyl alcohol. Benzylamine, serving as both the carbon source and nitrogen dopant, was introduced into the system in mist form from the nebulizer using a gas mixture of argon and 15% hydrogen. The patterned substrates were placed in the second furnace. The entire system was hermetically sealed with silicone paste and purged for 5 minutes with at a flow rate of 0.5 liter/min. The furnace temperature was ramped to 825 °C over 30 minutes. Upon reaching 825 °C, the nebulizer was activated, and the carrier gas flow (argon and 15% hydrogen) increased to 2.5 liter/min. After synthesis, the nebulizer was turned off, the gas flow rate was reduced back to 0.5 liter/min, and the furnace was cooled to 25°C, a process that typically took three hours.

**Fabrication of the AuNSt-AuCNT hybrid substrate**: The gold coating was deposited on the surfaces of CNT to prepare the AuCNT substrate by using a sputter coater instrument (CRESSINGTON sputter coater 108 auto, TED PELLA, INC., Redding, CA, USA). The current is 40 A with a deposition time of 120s. After that, 5 μL AuNSt solution was deposited on as-prepared AuCNT by drop casting and letting it dry in ambient air.

**Virus sample preparation**: Influenza A H1N1 (A/California/07/2009, A/Michigan/45/2015, A/Nebraska/14/2019), Influenza A H3N2 (A/Hawaii/47/2014, A/Arizona/45/2018, A/Delaware/39/2019, A/North Carolina/04/2016), and influenza B (B/Massachusetts/02/2012) viruses were propagated in Madin-Darby Canine Kidney-London (MDCK-London) cell cultures to viral titers with a TCID 50 (50% tissue culture infectious dose) of $10^6$-$10^7$. MDCK-London cells were maintained in Dulbecco's Modified Eagle's Medium (Invitrogen, Carlsbad, CA) supplemented with 10% fetal bovine serum and 1% penicillin-streptomycin, and incubated at 37 °C in a humidified $CO_2$ incubator. In contrast, avian infectious bronchitis virus (IBV, inoculated cell cultures were harvested when 70-100% cytopathic effects (CPE) were developed, generally in 48 – 72 h. ) was propagated in specific-pathogen-free (SPF) embryonated chicken eggs (ECE). A stock IBV reference strain or field isolate was diluted 1:5 in viral transport medium (VTM) and inoculated into 9-to-11-day-old ECE via the chorioallantoic cavity (0.2 mL per egg, using 3–5 eggs per sample). After incubation at 37 °C for 48–72 hours, the chorioallantoic fluid (CAF) containing IBV was harvested. For Raman data collection, 1 μL virus sample was deposited on as-prepared AuNSt-AuCNT hybrid substrate by drop casting and letting it dry in ambient air.

**FDTD simulation**: : Full-wave electromagnetic simulations were carried out by commercial software FDTD solutions 6.5 (Lumerical). The optical constants of Au were obtained from Palik[30]. The optical constants of CNT were from Ermolaev[31]. For total absorption, the simulation region was an 250 × 250 × 250 $nm^3$ cuboid space surrounded by Perfectly matched layer (PML) boundary conditions. A total field scattered field (TFSF) source was injected in the z direction and polarized in the x direction. The continuous-wave source (400-900 nm) was confined to only enters the range of a 180 × 180 × 180 $nm^3$ box. An absorption analysis group was placed inside of the TFSF source, with the range dimensions of 140 × 140 × 140 $nm^3$. The nanostar or nanosphere was placed inside the absorption analysis group at the center. For reflection simulation, the simulation region was 250 × 250 × 2200 $nm^3$. The plane wave source was injected in the z direction from z = 800 nm, and polarized in the x direction. The reflection was measured by a 2D z-normal frequency domain power monitor at z = 900nm. The nanostars were placed on Au/CNT substrates at z = 0 nm.

**Structure characterization**: *Scanning electron microscopy (SEM)*: SEM was conducted using Thermo Scientific Verios G4 operated at an accelerating voltage of 2 kV. The working distance was set at 3.6 mm, and the TLD detector was used for imaging. *Transmission electron microscopy (TEM)*: Samples were drop cast on to PELCO silicon nitride TEM support films. TEM images were acquired in an FEI Talos F200X TEM operated at 200 kV with an image acquisition of 1 seconds.

**EELS measurement**: The AuNSs-AuCNT hybrid structure was scratched from a $SiO_2$/Si substrate and dissolved into an isopropanol/water 1:1 mixed solvent with sonication. Then, 1 μL of the as-prepared solution was drop cast onto silicon nitride TEM support films. STEM EELS measurements were performed in a Titan$^3$ G2 60-300 S/TEM at 80 kV, utilizing a probe convergence semiangle of 30.2 mrad, and an EELS collection semiangle of 10.2 mrad. A monochromator was used to achieve an energy resolution of 0.200 eV, and EELS spectrum images were acquired with pixel size of 1.4 nm. To process the EELS datasets, the Hyperspy **(ref)** python package was used. To visualize the modes, the zero-loss peak was fit with a Voigt function to remove the zero-loss peak. Plasmon maps were obtained by creating an energy window (width = 0.1 eV) centered at the desired energy loss. A spectrum of an ROI is generated by summing the pixels in a 10 x 10 squared ROI.

**Optical characterization:** *Raman spectroscopy*: Raman measurements were performed using the Renishaw inVia confocal Raman system equipped with a 785 nm diode laser line and 1200 gr mm$^{-1}$ grating. The sample was focused on a 50X objective with the excitation laser power of approximately 1 mW and a spectral range from 500 to 2000 cm$^{-1}$. The typical exposure time for acquiring data was 10 s. *UV-vis spectroscopy*: The UV–vis spectra of AuNSt were measured using a LAMBDA 950 UV/Vis/NIR spectrometer over a 400–800 nm wavelength range. The AuNSt solution was placed in a quartz cuvette with a 10 mm optical path length. Water was used for baseline correction. *Reflection absorption spectroscopy*: Reflection absorption spectra were measured using a Horiba LabRAM Raman spectrometer. A 150/mm grating and a 100× objective lens with NA = 0.9 were used to collect reflected light over the 450–850 nm wavelength range.

**Principal Component Analysis and t-distributed stochastic neighbor embedding (t-SNE)**: Raman spectra were preprocessed using baseline correction, spike removal, and normalization to the maximum intensity of each spectrum. The preprocessed spectra were then input to PCA using the scikit-learn package in Python. The first three principal components were used for visualization in a 3D embedding space. For non-linear dimensionality reduction, t-SNE was performed using scikit-learn with the following parameters: perplexity = 200, learning rate = 200, n_iter = 1000. Spectra were visualized in 3D embedding spaces to assess clustering and separation among different virus types, subtypes, and strains.

**Machine Learning Classification**: For supervised classification of Raman spectra, three machine learning models were implemented: a one-dimensional convolutional neural network (1D CNN), support vector machine (SVM), and XGBoost. All analyses were conducted using Python. Prior to classification, all spectra were preprocessed using a standardized pipeline consisting of baseline correction, spike removal, and normalization by the maximum intensity of each spectrum. To ensure balanced class representation for each classification task (e.g., virus type, subtype, or strain), the dataset was balanced by random sampling. The processed data were then partitioned into training and validation sets using 5-fold stratified cross-validation. Our 1D-CNN model consists of 5 sequential Conv1D convolution layers, followed by a fully connected linear layer. The primary objective of our 1D-CNN design is to maximize the receptive field while maintaining a compact and lightweight architecture to reduce the risk of overfitting. The first four Conv1D convolution layers use a relatively large kernel size of 11 and a stride of 2, enabling the model to efficiently capture long-range dependencies and global spectral features. Each of these layers is followed by batch normalization, a ReLU activation function, and max pooling, which rapidly reduces the spatial resolution and parameter count. The fifth and final Conv1D convolution layer uses a smaller kernel size of 3 to allow for more compact and fine-grained feature representation. This layer is followed by adaptive average pooling, flattening, and a fully connected linear layer. To further mitigate overfitting, dropout is applied after the fully connected layer during training. The network was trained using the Adam optimizer with a learning rate of 0.01, batch size of 512, and 100 training epochs. Cross-entropy loss was used as the objective function, and learning rate scheduling was applied to improve convergence. Model performance was monitored using validation accuracy. For comparison, SVM classifiers with a linear kernel ($C = 1$) and XGBoost classifiers with default parameters were trained on the same datasets. All models were evaluated using 5-fold cross-validation, and their classification performance was assessed using precision, recall, and F1-score. To interpret the models' predictions, feature importance was estimated for each classifier to identify the most influential Raman shifts contributing to classification. For the CNN, a full-gradient method was used. For the SVM, feature importance was derived from the absolute values of the learned coefficients. For XGBoost, feature importance scores were computed based on the gain and frequency of feature usage during tree construction. These feature importances were visualized as a function of Raman wavenumber, enabling interpretation of the spectral regions most relevant to virus identification. All code was implemented using PyTorch for deep learning, scikit-learn for SVM and cross-validation, and XGBoost for gradient boosting.

**Acknowledgement**

The authors thank A.P. Schmitt and P.T. Schmitt for the help with viral sample preparation. This work was supported by the National Science Foundation's Growing Convergence Research Big Idea (under Grant ECCS-1934977). This work was funded in part by the Division of Intramural research (DIR) at te National Institute of Allergy and Infectious Diseases, National Institutes of Health (NIAID/NIH). The contributions of the NIH authors were made as part of their official duties as NIH federal employees, are in compliance with agency policy requirements, and are considered Works of the United States Government. However, the findings and conclusions presented in this paper are those of the authors and do not necessarily reflect the views of the NIH or the U.S. Department of Health and Human Services. We also thank the National Science Foundation (NSF) under DMR-1420620 and DMR-2011839 through the Penn State MRSEC−Center for Nanoscale Science for partial financial support. Z.W., X.W., and S.H. also acknowledge the support from NSF (ECCS-2246564, ECCS-1943895, and ECCS-2230400) and the Welch Foundation (C-2144).

**Author contributions**

N.Z. and Z.W. contributed equally to this work. N.Z and M.T. conceived the project. S.H. and M.T. supervised the project. N.Z. conducted the sample preparation and spectroscopy measurement. Z.W., N.Z., S.L., and S. H. performed machine learning analysis with the help of J.Y. and S.X.H. X.W. and S. H. contributed to the FDTD simulation. G.A.V-L and K.D.H conducted EELS measurements under the supervision of D.R.H. P.P.V. and D.J.A. synthesized gold nanostars under the supervision of L.M.L-M. N. Z., D.E.S, N.P., E.D., and A.J.S. contributed to the structural characterization. R.R, H.L., and E.G. contributed to virus sample preparation. N.P.,Y-T.Y., and J.A.M. contributed to synthesis of carbon nanotubes. N.Z. and Z.W. performed the analysis of the data and wrote the manuscript with the input from all coauthors. All the authors reviewed and revised the paper.